\documentclass[%
 reprint,
 superscriptaddress,
 twocolumn, 
 amsmath,amssymb,
 aps,
]{revtex4-2}
\usepackage{todonotes}
\usepackage[colorlinks = true,
            linkcolor = blue,
            urlcolor  = blue,
            citecolor = blue,
            anchorcolor = blue]{hyperref}
\usepackage[normalem]{ulem}
\usepackage{graphicx}
\usepackage{dcolumn}
\usepackage{bm}
\usepackage{braket}
\usepackage{glossaries}
\usepackage{color}
\definecolor{darkred}{rgb}{0.80,0,0}
\definecolor{blood}{rgb}{0.50,0,0}
\definecolor{brightred}{rgb}{1,0,0}
\definecolor{orange}{rgb}{1,0.3,0}
\definecolor{bluegreen}{rgb}{0,0.5,0.5}
\definecolor{lightblue}{rgb}{0,0.5,0.8}
\definecolor{darkgreen}{rgb}{0,0.5,0}
\definecolor{green}{rgb}{0,0.70,0}
\definecolor{darkblue}{rgb}{0,0,0.80}
\definecolor{magenta}{rgb}{1,0,1}
\definecolor{softmagenta}{rgb}{0.85,0.1,0.6}
\definecolor{mauve}{rgb}{0.6,0.1,1}
\definecolor{white}{rgb}{1,1,1}
\definecolor{black}{rgb}{0,0,0}

\newacronym{CDW}{CDW}{charge-density-wave}
\newacronym{QMC}{QMC}{quantum monte carlo}
\newacronym{DFPT}{DFPT}{density functional perturbation theory}
\newacronym{DFT}{DFT}{density functional theory}

\usepackage{ulem}

\makeatletter
\newsavebox{\@brx}
\newcommand{\llangle}[1][]{\savebox{\@brx}{\(\m@th{#1\langle}\)}%
  \mathopen{\copy\@brx\kern-0.5\wd\@brx\usebox{\@brx}}}
\newcommand{\rrangle}[1][]{\savebox{\@brx}{\(\m@th{#1\rangle}\)}%
  \mathclose{\copy\@brx\kern-0.5\wd\@brx\usebox{\@brx}}}
\makeatother

\newcommand{\COMMENTED}[1]{}

\begin{document}

 \title{Finite-momentum pairing and magnetic halos in a spin-imbalanced
Holstein model}

 \author{Chunhan Feng}
\affiliation{Max Planck Institute for the Physics of Complex Systems, Nöthnitzer Straße 38, 01187 Dresden, Germany}
  \author{George Batrouni}
 \affiliation{Universit\'e C\^ote d'Azur, CNRS, Institut de Physique de Nice (INPHYNI), 06000 Nice, France}
\author{Richard Scalettar}
\affiliation{Department of Physics and Astronomy, University of California, Davis,
  California 95616, USA}

\date{\today}

\begin{abstract}

Fulde-Ferrell-Larkin-Ovchinnikov (FFLO) superconductivity, Cooper pairing at finite-momentum driven by spin imbalance, remains difficult to study numerically in two dimensions. The fermion sign problem in the spin-imbalanced attractive Hubbard model severely limits quantum Monte Carlo access to the low temperatures where pairing correlations are most pronounced. We show that the Holstein model, in which the effective attraction is retarded by phonon dynamics, strongly suppresses sign fluctuations under spin imbalance, enabling unbiased determinant quantum Monte Carlo simulations deep in this regime. With increasing magnetization, zero-momentum pairing evolves into finite-momentum FFLO pairing, evidenced by oscillating real-space pair correlations and a shift of the pairing susceptibility peak to finite $\vec{Q}$. The finite-momentum susceptibility substantially exceeds its zero-momentum value. At high density $n \sim 0.875$, near magnetization $m \approx 1-n$ where the majority-spin density approaches half filling, FFLO pairing coexists with checkerboard charge-density-wave order at short range, while staggered longitudinal spin correlations emerge as a magnetic halo at larger distances. These spatial textures provide experimentally accessible signatures of finite-momentum pairing and motivate cold-atom realizations of FFLO physics.

\end{abstract}

\maketitle

\noindent
\textit{I.~Introduction}.~
In conventional spin singlet superconductivity, spin-up fermions of momentum $\vec k$
pair with spin-down electrons of momentum $-\vec k$, leading to an order parameter
$\langle \hat \Delta^{\dagger}(\vec r \,) \rangle \equiv \langle \hat c^{\dagger}_{\vec r \uparrow} \,
\hat c^{\dagger}_{\vec r \downarrow} \rangle = \Delta_{0}$ which is spatially uniform.  
The formation of Cooper pairs with momentum $\vec k + (-\vec k) = \vec 0$ is especially
straightforward in the case of equal populations of spin $\uparrow$ and
spin $\downarrow$ owing to their sharing of a common Fermi surface.
However,  an applied (Zeeman) magnetic field $B$ splits the band structure
of the two spin species, leading to distinct Fermi surfaces.
A possible consequence is 
pairing of a spin-up fermion of momentum $\vec k $ with
a spin-down fermion of momentum $-\vec k \pm \vec Q$. Such Cooper pairing
at non-zero momentum $\pm \vec Q$, is termed Fulde-Ferrell-Larkin-Ovchinnikov (FFLO) superconductivity~\cite{fulde1964,larkin1964}.  The resulting pairing correlations are spatially modulated, with $\vec Q$ set by the mismatch between the spin-split Fermi surfaces.

Despite its conceptual clarity,
FFLO superconductivity is difficult to achieve experimentally.
Impurity scattering tends to wash out the oscillations of
the order parameter unless the mean free path exceeds the pairing coherence length.
Similarly, low temperatures are required to prevent thermal smearing of the matching of
the two spin-split Fermi surfaces.
As a consequence, the experimental observation of an FFLO phase has been challenging.
Leading candidate materials include heavy fermion systems like
CeCoIn$_{5}$ \cite{matsuda2007} and, more recently,
ultracold atomic gases~\cite{kinnunen2006,Mitra2016,Chevy2006a,Hartke2026}.

FFLO correlations are well established theoretically in one dimension, with analytical and numerical studies providing extensive evidence for their finite-momentum pairing character and spatial structure \cite{cheng2018,tezuka2008,luscher2008,tezuka2010,sano2013,batrouni2008,rizzi2008,casula2008,batrouni2009,machida2006,machida2008}. One dimension is, however, special: superconducting correlations  do not develop true long-range order even at zero temperature, while the Fermi surface consists only of discrete Fermi points, making the momentum matching required for finite-$\vec{Q}$ pairing comparatively favorable.

In two dimensions, however, the problem is considerably more challenging. Mean-field approaches have established the basic FFLO phenomenology, while numerical studies -- including DMRG on quasi-1D systems, determinant quantum Monte Carlo (DQMC), diagrammatic Monte Carlo, and constrained-path auxiliary-field quantum Monte Carlo -- have provided evidence for finite-momentum pairing across different parameter regimes \cite{moreo2007,feiguin2009,loh2010, gukelberger2016,wolak2012,feng2026,Qu2026,Karlsson2026}. Yet an unbiased, low-temperature understanding of FFLO in two dimensions remains challenging. For the spin-imbalanced attractive Hubbard model, the unequal fermion determinants cause a severe sign problem, with previous DQMC studies accessing low temperatures only in the low-density regime \cite{wolak2012}.

The Holstein model provides a distinct setting in which attraction arises from a retarded electron-phonon coupling, continuously connecting to the attractive Hubbard model in the anti-adiabatic limit. Away from this limit, its dynamical phonon field strongly suppresses sign fluctuations under spin imbalance, enabling unbiased DQMC access deep in the low-temperature regime. Our results here establish three key findings.
(i) quantifying the sign problem suppression by phonon dynamics;
(ii) Increasing magnetization drives an evolution from zero- to finite-momentum pairing, with the finite-$\vec{Q}$ susceptibility exceeding the zero-momentum value by up to 70\%. (iii) At high density, near $m\approx1-n$, FFLO pairing dominates at short range with checkerboard charge-density-wave (CDW), while staggered longitudinal spin correlations emerge as a halo at larger distances.
These results open a new avenue for studying intertwined FFLO pairing, charge, and spin correlations in two dimensions, while identifying experimentally accessible signatures of FFLO correlations and motivating cold-atom realizations.

\vskip0.07in \noindent
\textit{II.~Hamiltonian and Methodology.} 
We consider the two-dimensional Holstein model on a square lattice in the presence of a 
uniform Zeeman magnetic field. The Hamiltonian is given by
\begin{equation}
\begin{aligned}
\hat{\cal H} ={}& -t \sum_{\langle ij\rangle,\sigma}
\left(
\hat{c}_{i\sigma}^{\dagger}\hat{c}_{j\sigma}
+\mathrm{H.c.}
\right)
-\mu\sum_{i,\sigma}\hat{n}_{i\sigma}
-2B\sum_i \hat{S}_i^z \\
&+\sum_i
\left(
\frac{\hat{P}_i^2}{2M}
+\frac{M\omega_0^2}{2}\hat{X}_i^2
\right)
+\lambda\sum_i
\hat{X}_i \hat{n}_i.
\end{aligned}
\label{eq:holsteinham}
\end{equation}
Here, $\hat{c}_{i\sigma}^{\dagger}$ ($\hat{c}_{i\sigma}$) creates (annihilates) an electron with spin $\sigma=\uparrow,\downarrow$ at site $i$, and
$\hat{n}_{i\sigma}=\hat{c}_{i\sigma}^{\dagger}\hat{c}_{i\sigma}^{\phantom{\dagger}}$ with
$\hat{n}_i=\hat{n}_{i\uparrow}+\hat{n}_{i\downarrow}$.
The first term describes nearest-neighbor hopping with amplitude $t$ on the square lattice. 
The chemical potential $\mu$ controls the global filling (both spin species)
while a Zeeman field $B$ couples to
 $\hat{S}_i^z=\frac{1}{2}\left(\hat{n}_{i\uparrow}-\hat{n}_{i\downarrow}\right)$. 
The second line represents the kinetic and potential
 energy of local Einstein phonons with frequency $\omega_0$
 and a local electron-phonon interaction with coupling 
strength $\lambda$. Throughout this work, we set $t=1$ as the unit of energy, and use $M=1$, $\omega_0=1$ and $\lambda=1.4$. $n=\langle n_{i\uparrow} + n_{i\downarrow} \rangle$ and $m=\langle n_{i\uparrow} - n_{i\downarrow} \rangle$ denote the density and magnetization of the system respectively. Periodic boundary conditions are used in the simulation.

 \begin{figure}[t!]
  \includegraphics[width=1\columnwidth]{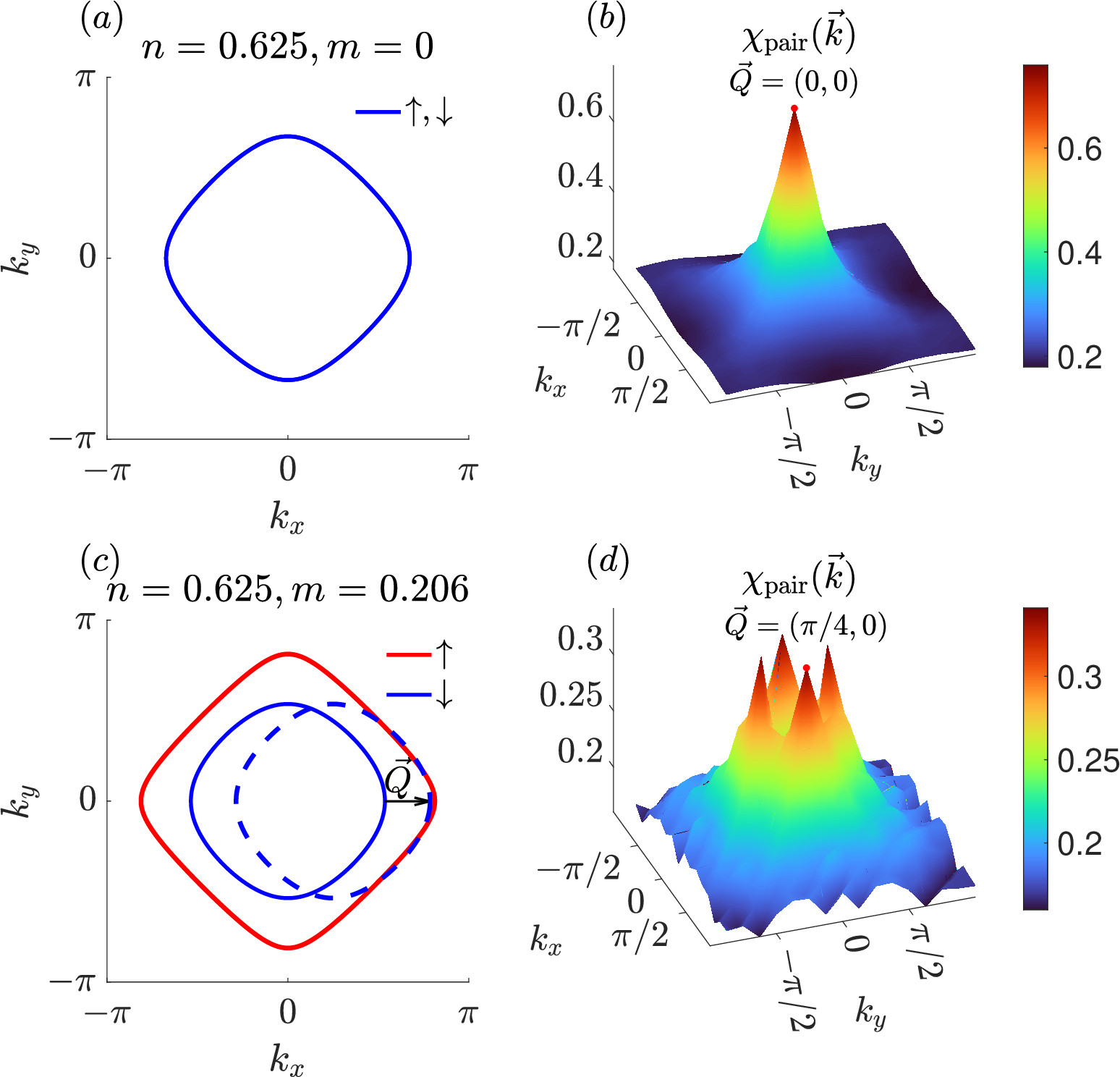}
 \caption{\textbf{Fermi-surface matching and finite-momentum pairing.}
 Non-interacting Fermi surfaces at $n=0.625$ for (a) zero magnetization, $m=0$, and (c) finite magnetization, $m=0.206$. Panels (b) and (d) show the corresponding momentum-dependent pair susceptibilities $\chi_{\rm pair}(\mathbf k)$.  At $m=0.206$ (d), $\chi_{\rm pair}(\vec{k})$ is maximum at $\vec{Q}=(\pi/4,0)$ and its symmetry-equivalent momenta. In (c), dashed line denotes the spin-down Fermi surface shifted by $\vec{Q}=(\pi/4,0)$.
 Calculations are done on a $16 \times 16$ lattice at $\beta=16$.
 }
  \label{fig:volcano}
  \end{figure}

We study the Hamiltonian 
using the DQMC method
\cite{blankenbecler1981,white89,assaad02,gubernatis16}.
The partition function is written as a path integral over the phonon field
with the quadratic fermion degrees of freedom integrated out.  The resulting
configurational weight consists of two fermion determinants, one for spin-up and one for spin-down,
which depend on the fermion hopping parameters and the electron-phonon coupling,
as well as a `bosonic' contribution coming from the purely phonon pieces of the Hamiltonian.
For $B=0$ the up and down spin determinants are identical, and the weight is positive.  Thus,
as with the attractive Hubbard model, simulations can be performed down to very low temperature $T$ with no sign problem.  
Measurements are obtained by combining appropriate products of the fermion Green's functions,
the inverse of the matrices whose determinants enter the weight.
There is an extensive literature on QMC approaches to
the spin-balanced Holstein Hamiltonian\cite{scalettar89,marsiglio90,li2015-1,Zhang2019,Chen2019,Feng2020,Feng2020b,Bradley2021,Yao2021,Nosarzewski2021,Zhang2022,Araujo2022,Liu2026,Faundez2026,Neuhaus2026,Liu2026}.

However, this symmetry is broken for $B \neq 0$ and a sign problem ensues.
For the attractive Hubbard model, this limits accessible temperatures to $T/t \gtrsim 0.3$, and the densities to very low values,
depending on the precise Hamiltonian parameters and spatial lattice size\cite{wolak2012}. The 
Holstein case is better behaved because, in  the construction of the path integral, 
the phonon kinetic energy $\hat P^2 / 2M$ yields a coupling between phonon coordinates on
adjacent time slices \cite{creutz1981statistical}.  The resulting suppression of imaginary time fluctuations mitigates negative determinants.  Although negative determinants still occur in the parameter regime studied here, the average sign remains above 0.9 (see SM), allowing reliable results to be obtained with $10^4$--$10^5$ Monte Carlo sweeps.

To characterize the magnetic, charge, and superconducting properties of the system, we measure the corresponding real-space correlations,
$C_{\alpha}(\vec{r})=\langle (\hat S_i^{\alpha} - \langle \hat S_i^{\alpha}\rangle) (\hat S_{i+\vec{r}}^{\alpha} - \langle S_{i+\vec{r}}^{\alpha} \rangle )\rangle,\quad
C_{n}(\vec{r})=\langle (\hat{n}_i- n)(\hat{n}_{i+\vec{r}}- n)\rangle,\quad
C_{p}(\vec{r})=\langle \Delta_i^\dagger\Delta_{i+\vec{r}} +h.c. \rangle $,
where $\alpha=x,z$ denotes the spin component and $\hat \Delta^{\dagger}_{i} = 
\hat c^{\dagger}_{i\uparrow} \hat c^{\dagger}_{i\downarrow}$ is the pairing order parameter. Their Fourier transforms define the corresponding structure factors,
$S_{\mu}(\vec k)=\sum_{\vec{r}}e^{i\vec k\cdot\vec{r}}C_{\mu}(\vec{r})$, where $\mu=\alpha,n,p$.
Integration over imaginary time yields the susceptibilities associated with these observables.
 Of particular interest is the
pairing susceptibility, 
$\chi_{\rm pair}(\vec Q) = \int_{0}^{\beta} d\tau \, \sum_{\vec r} e^{i \vec Q \cdot \vec r} \,
\langle \,  e^{\tau \hat H} \hat \Delta^{\phantom{\dagger}}_{i+r} e^{-\tau \hat H} \, \hat \Delta^{\dagger}_{i} +h.c. \, \rangle$,
where $\beta$ is the inverse temperature.
The imaginary time integration allows greater sensitivity to superconductivity.

 \begin{figure}[t!]
  \includegraphics[width=1\columnwidth]{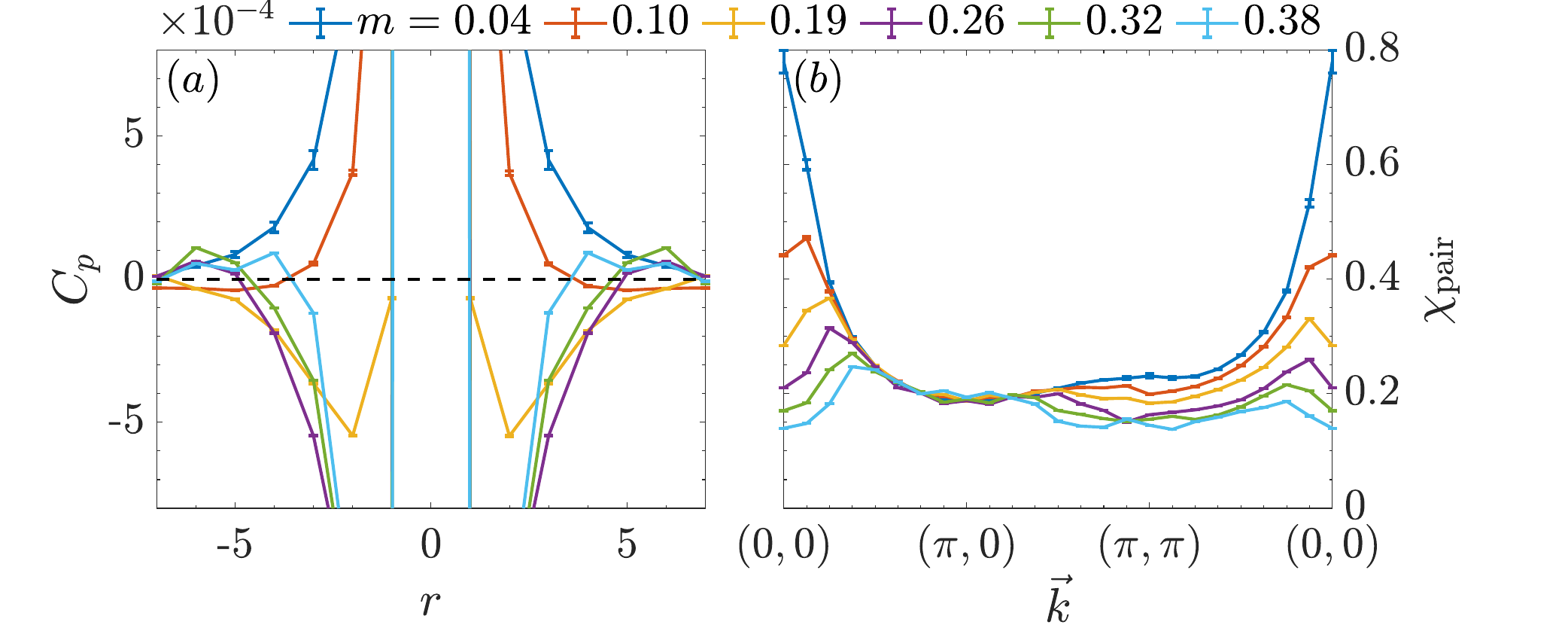}
 \caption{\textbf{Signatures of FFLO superconductivity.} (a) Real space pairing correlation $C_p(\vec r\,)$ along the diagonal line cut $\vec r = (r,r)$ at 
 different magnetizations $m$. (b) Pair susceptibility $\chi_{\rm pair}$ along momentum line-cut $\vec{k}=(0,0)\rightarrow (\pi,0) \rightarrow (\pi,\pi) \rightarrow (0,0)$.
  Parameters are the same as in Fig.~\ref{fig:lambda14_den0875_Sx_pairing}. Calculations are performed on a $16 \times 16$ lattice with density $n=0.875$ at $\beta=16$.
 The large, positive $r=0$ value is suppressed to enable the plot to focus
 on the smaller off-site correlations.
 }
  \label{fig:pair_r_k_lambda14_den0875}
  \end{figure}

\vskip0.07in \noindent
\textit{III.~Results.} We first establish the connection between Fermi-surface matching and finite-momentum pairing in Fig.~\ref{fig:volcano}. 
For the spin-balanced system at density $n=0.625$ and magnetization $m=0$, the spin-up and spin-down Fermi surfaces coincide [Fig.~\ref{fig:volcano}(a)], allowing fermions at momenta $\vec{k}$ and $-\vec{k}$ to pair with zero center-of-mass momentum. 
Accordingly, the pair susceptibility $\chi_{\rm pair}(\vec Q)$ is maximal at $\vec Q=(0,0)$ [Fig.~\ref{fig:volcano}(b)]. 
Upon spin imbalance, $m=0.206$, the magnetic field separates the two Fermi surfaces [Fig.~\ref{fig:volcano}(c)]. 
Pairing at finite center-of-mass momentum $\pm \vec Q$ can then improve the Fermi-surface matching, with $\vec{k}$ paired with $-\vec{k}\pm\vec Q$. 
The pair susceptibility correspondingly develops a maximum at $\vec Q=(\pi/4,0)$ [Fig.~\ref{fig:volcano}(d)], where $\chi_{\rm pair}(\vec Q)/\chi_{\rm pair}(0)=1.54$, demonstrating a substantial enhancement of finite-momentum pairing.

We next track the evolution of the finite-momentum pairing with magnetization in real and momentum space. 
We fix the density at $n=0.875$ and use it as a representative case. 
Figure~\ref{fig:pair_r_k_lambda14_den0875}(a) shows the pairing correlation $C_p(\vec r)$ along the diagonal direction, $\vec r=(r,r)$. 
This choice is motivated by the fact that the FFLO node in the attractive Hubbard model, and the $\pi$-phase shift of stripe order in the doped repulsive Hubbard model, first emerge along the diagonal direction \cite{feng2026}. 
At weak magnetization, $m=0.04$, $C_p(\vec r)$ remains positive throughout the accessible system size. 
Here the Fermi-surface mismatch $\Delta k$ is smaller than the minimum momentum resolution, $2\pi/16$, of the $16\times16$ lattice, yielding an FFLO wavelength too long for its first node, expected at $r\sim\lambda_{\rm FFLO}/4$, to fall within the accessible range $r\le L/2$. 
Upon increasing the magnetization to $m=0.10$, a first node appears at $r\sim3$--$4$. 
The node subsequently moves toward shorter distances $r\sim1$--$2$ with increasing $m$, reflecting the decreasing FFLO wavelength. 
At $m=0.26$, a second node becomes visible, with $C_p(r)$ recovering to positive values near $r\sim5$.

The momentum-space pair susceptibility directly identifies the corresponding pairing momentum. 
Figure~\ref{fig:pair_r_k_lambda14_den0875}(b) shows $\chi_{\rm pair}(\vec Q)$ along the path $(0,0)\rightarrow(\pi,0)\rightarrow(\pi,\pi)\rightarrow(0,0)$. 
With increasing magnetization, the dominant susceptibility shifts from $\vec Q=(0,0)$ to finite momentum, with pronounced maxima along both the $(1,0)$ and $(1,1)$ directions, consistent with previous results for the attractive Hubbard model~\cite{feng2026}. 
For $m=0.38$, for example, $\chi_{\rm pair}(3\pi/8,0)$ exceeds $\chi_{\rm pair}(0,0)$ by approximately $70\%$. 
This enhancement is substantially larger than the less than $10\%$ enhancement of the equal-time pairing correlator reported for the two-dimensional attractive Hubbard model at $U=-3.5$~\cite{wolak2012}\footnote{This was at polarization 
$P=0.6$ and $\beta=10$, 
on a  $16 \times 16$ lattice.  In order to reach this
large $\beta$ and still have a manageable sign problem, the system had to be made very dilute: 
$\rho \sim 0.3$.}.

 \begin{figure}[t!]
  \includegraphics[width=1\columnwidth]{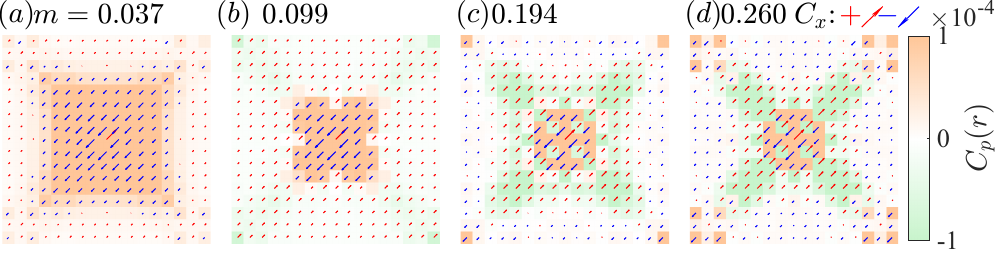}
  \includegraphics[width=1\columnwidth]{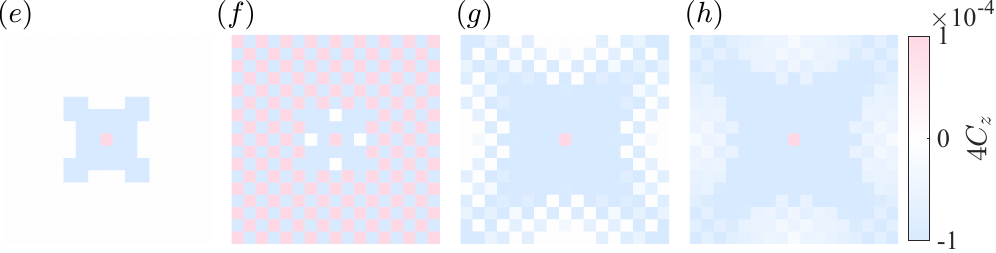}
    \includegraphics[width=1\columnwidth]{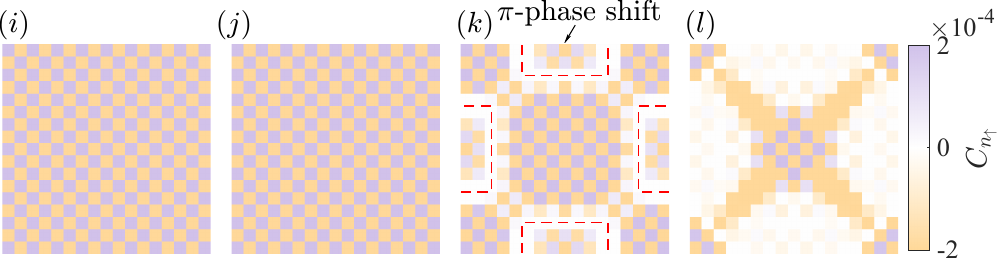}
\caption{\textbf{Intertwined Orders.} (a-d): Orange and green denote positive and negative pairing correlations $C_p$, respectively. Red and blue arrows denote positive and negative spin-X correlations $C_x$. (e-h): Pink and blue denote positive and negative spin-Z correlations $C_z$. (i-l) purple and yellow denote positive and negative spin-up density correlations $C_{n_{\uparrow}}$. Red dashed lines enclose regions where CDW exhibits a $\pi$ phase shift. All correlations are measured relative to the central site, with arrow sizes scaling with the corresponding correlation strength. FFLO pairing and CDW order dominate the central region, while for $m\sim1-n$, staggered magnetic correlations emerge in the pairing-suppressed outer region. Calculations are performed on a $16\times16$ lattice, $n=0.875$ at $\beta=16$.
}
  \label{fig:lambda14_den0875_Sx_pairing}
  \end{figure}

\begin{figure*}[t!]
 \includegraphics[width=2\columnwidth]{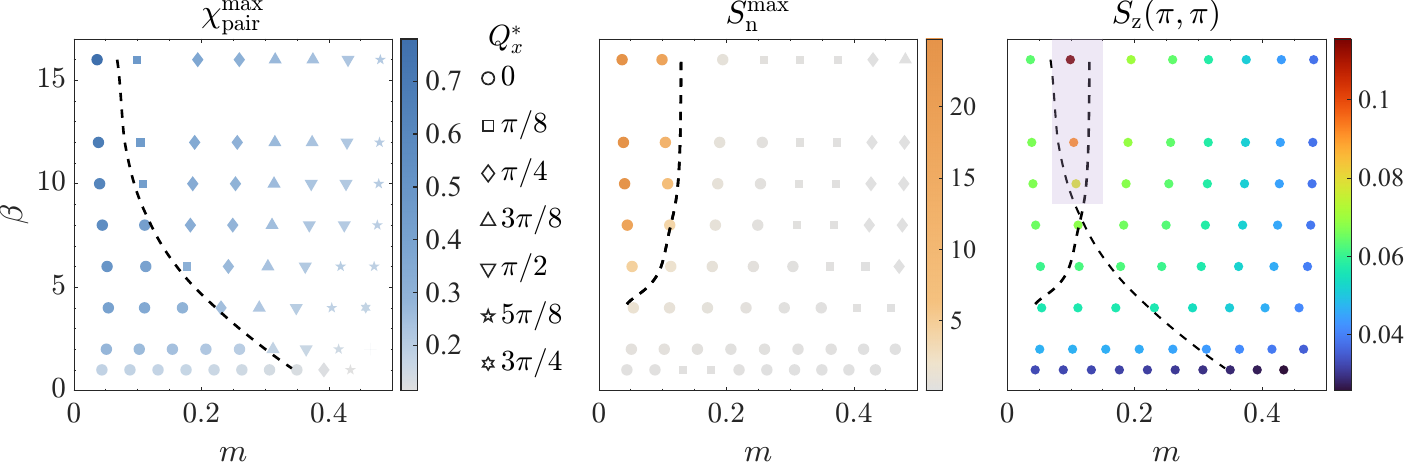}
\caption{\textbf{Evolution of the FFLO halo in the magnetization $m$--inverse-temperature $\beta$ plane.} Maximum (a) pair susceptibility, (b) CDW structure factor, and (c) spin-Z correlation $S_z(\pi,\pi)$ as functions of $m$ and $\beta$. Symbols indicate the shift of the peak wave vector $(\Delta Q_x^*,0)$ away from $(0,0)$ in (a) and $(\pi,\pi)$ in (b). Dashed lines are guides to the eye. The region to the right of the dashed line in (a) indicates finite-momentum pairing, while that to the left of the dashed line in (b) indicates strong CDW. The dashed lines in (c) reproduce those in (a) and (b). The purple-shaded intermediate regime denotes the region where a halo of staggered spin-Z correlations is observed. Parameters are the same as in Fig.~\ref{fig:lambda14_den0875_Sx_pairing}.
 }
  \label{fig:phase_diagram}
  \end{figure*}


Having established the presence of finite-momentum FFLO pairing, we now examine its interplay with charge and spin correlations in real space. 
Figure~\ref{fig:lambda14_den0875_Sx_pairing} summarizes the pair, spin, and charge correlations across magnetization at fixed density $n=0.875$ and $\beta=16$.
At weak magnetization, $m=0.037$, the pairing correlations remain positive throughout the system [Fig.~\ref{fig:lambda14_den0875_Sx_pairing}(a)], consistent with conventional zero-momentum pairing. The transverse spin correlations $C_x(\vec r)$ are predominantly negative away from the origin, while the charge correlations exhibit a pronounced checkerboard pattern, reflecting the strong charge-ordering tendency induced by the effective attractive interaction of the Holstein coupling at this relatively high density.
The connected longitudinal spin correlations
$C_z(\vec r)$
are negative near the origin but rapidly decay toward zero at larger separations, indicating only short-range longitudinal spin correlations.

For larger magnetizations in Fig.~\ref{fig:lambda14_den0875_Sx_pairing}, FFLO superconductivity emerges. Most prominently, the pair correlations $C_p$ develop a spatial modulation with sign changes, which is accompanied by corresponding sign changes in the spin-X correlations $C_x$. A useful analogy is provided by the repulsive Hubbard model, where spatial modulation of the spin stripe is typically accompanied by charge-density modulations, with the charge channel being the particle-hole counterpart of the spin-Z channel. Here, the pairing channel and the transverse spin-X/Y channel play an analogous role, with the latter being the particle-hole counterpart of the former. This provides a natural way to understand the corresponding modulation of the transverse spin correlations observed alongside the FFLO pairing. A similar correspondence between FFLO pairing and transverse spin correlations has also been observed in the spin-imbalanced attractive Hubbard model~\cite{feng2026}.

More strikingly, at intermediate magnetization $m=0.099$ and $0.194$, where the majority-spin density is close to half filling, the correlations develop a pronounced spatial structure around the reference site. Pairing and charge correlations dominate at short distances, while staggered connected longitudinal spin correlations form a magnetic halo at larger distances [Figs.~\ref{fig:lambda14_den0875_Sx_pairing}(b--c), (f--g), and (j--k)]. At $m=0.099$, the halo exhibits an approximately positive-negative alternation in $C_z$, whereas at $m=0.194$ the pattern evolves toward a nearly zero-negative alternation. At larger magnetization, $m=0.260$, the staggered structure disappears. The halo is most pronounced near $m\sim1-n\approx0.125$, where $n_\uparrow\sim0.5$. This can be understood from the interplay between pairing and excess majority fermions: the paired particles dominate the short-distance correlations, while the remaining majority fermions are subject to Pauli exclusion and, near half filling, a perfectly nested Fermi surface at $\vec Q=(\pi,\pi)$, favoring staggered spin correlations.

The charge correlations undergo a related but distinct evolution. Since the magnetic halo 
reflects the spatial distribution of excess majority fermions, we track the majority-spin 
density $C_{n\uparrow}$, which also captures the overall charge-ordering tendency.
At $m=0.099$, the checkerboard CDW persists across the system, although its amplitude is reduced relative to the weakly magnetized case. 
At $m=0.194$, where $n_\uparrow\sim0.535>0.5$, the CDW develops a $\pi$ phase shift between the central and outer regions.
A related $\pi$ phase shift is expected in the anti-adiabatic limit at half filling for a spin-imbalanced system with $n_\uparrow>0.5$ and $n_\downarrow<0.5$, where the pseudospin SU(2) symmetry relates the pairing and CDW channels \cite{moreo2007}. 
In that limit, the two correlations share the same characteristic wavelength and node positions. 
The doped Holstein model considered here explicitly breaks this symmetry, and the pairing and charge correlations consequently could develop distinct spatial structures.
Indeed, the CDW $\pi$-phase shift occurs only at a longer distance, $r\sim7$ [Fig.~\ref{fig:lambda14_den0875_Sx_pairing}(k)], compared with the first pairing node at $r\sim3$ [Fig.~\ref{fig:lambda14_den0875_Sx_pairing}(c)]. 
This CDW $\pi$-phase shift at $r\sim7$ also approximately coincides with the boundary between the central pairing/CDW region and the outer region dominated by staggered spin-Z correlations.

\begin{figure}[t]
  \includegraphics[width=1\columnwidth]{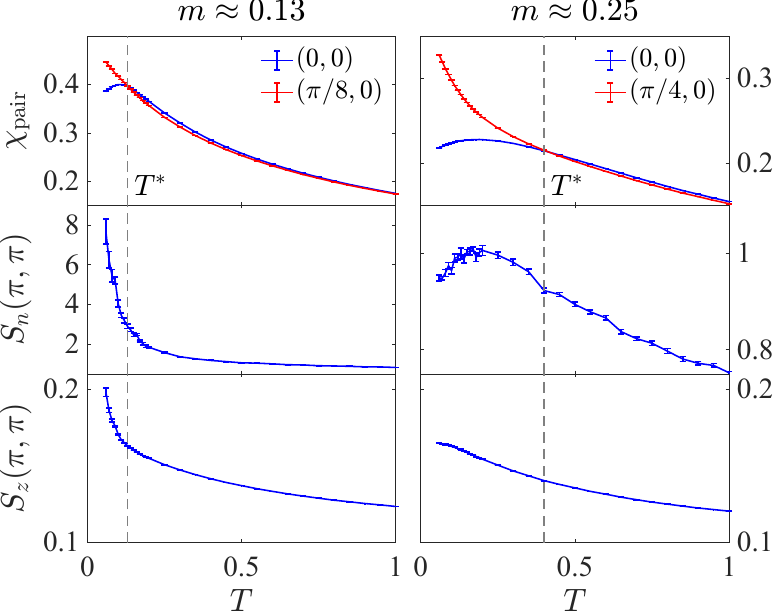}
 \caption{\textbf{Pair susceptibility and charge and spin-Z structure factors.} As functions of temperature $T$ for two magnetizations, $m\approx0.13$ and $0.25$. The dashed line marks the temperature $T^*$ below which the finite-momentum pair susceptibility exceeds its zero-momentum value, indicating the onset of FFLO pairing. At $m\approx0.13$, where the FFLO with a magnetic halo is observed, both $S_n$ and $S_z$ exhibit a sharp increase just below $T^*$. In contrast, at $m=0.25$, where no FFLO halo is observed, both structure factors remain substantially smaller and show no comparable increase.
 }
  \label{fig:lambda14_den0875_fix_m_vs_T}
  \end{figure}

The phase diagram in Fig.~\ref{fig:phase_diagram} summarizes the evolution of pairing, charge, and spin patterns over the $m$--$\beta$ plane. The pairing susceptibility [Fig.~\ref{fig:phase_diagram}(a)] shows a systematic shift of its maximum away from $\vec Q=(0,0)$ with increasing magnetization. At fixed temperature, the characteristic pairing momentum increases with $m$, consistent with the increasing Fermi-surface mismatch discussed earlier. Finite-momentum pairing also extends to higher temperatures with increasing magnetization. To the right of the dashed line, finite-momentum pairing becomes dominant. The charge structure factor [Fig.~\ref{fig:phase_diagram}(b)] remains strongly peaked at $(\pi,\pi)$ over the low-magnetization regime. At larger magnetization, the emergence of a spatial $\pi$ phase shift checkerboard CDW causes the dominant charge wave vector to move away from $(\pi,\pi)$. Because these charge correlations decay relatively rapidly, however, the displaced peak remains modest in magnitude. The longitudinal spin structure factor $S_z(\pi,\pi)$ [Fig.~\ref{fig:phase_diagram}(c)] is strongly enhanced around $m\sim1-n$, coinciding with the regime where staggered spin-Z correlations emerge in the outer region of the system. The shaded region traces the parameter range where the FFLO coexists with magnetic correlations, centered around $m\sim 1-n$.

Finally, we examine the temperature evolution of the pairing, charge, and spin correlations in Fig.~\ref{fig:lambda14_den0875_fix_m_vs_T}. We compare $m\sim0.13$, where FFLO coexists with a magnetic halo, with $m\sim0.25$, where no halo is present. At $m\sim0.13$, the finite-momentum pair susceptibility at $\vec Q=(\pi/8,0)$ overtakes the zero-momentum susceptibility at $T^*\sim0.1$. Just below this temperature, both the charge structure factor $S_n(\pi,\pi)$ and the longitudinal spin structure factor $S_z(\pi,\pi)$ increase sharply, coinciding with the emergence of the magnetic halo. In contrast, at $m\sim0.25$, finite-momentum pairing develops at a substantially higher temperature, $T^*\sim0.4$. Below $T^*$, $S_n(\pi,\pi)$ shows a weaker enhancement, while the zero-momentum pairing susceptibility exhibits a similar behavior, with both beginning to decrease upon further cooling.
The longitudinal spin response, however, remains comparatively weak, consistent with the absence of a magnetic halo. 
  
\vskip0.07in \noindent
\textit{IV.~Conclusions.} 
We have characterized finite-momentum pairing in the spin-imbalanced two-dimensional Holstein model using determinant quantum Monte Carlo. With increasing magnetization, conventional zero-momentum pairing evolves into pronounced finite-$Q$ pairing, accompanied by a reconstruction of the charge and spin correlations. 
Here, finite-momentum pairing emerges from a retarded phonon-mediated interaction, providing a distinct microscopic setting for FFLO physics beyond the conventional attractive Hubbard description. Remarkably, we find this regime remains accessible with only a very mild sign problem in the parameter range studied here, providing an unbiased route to low-temperature FFLO physics in two dimensions.

Our results further reveal that FFLO pairing is intertwined with spatially structured charge and spin correlations, forming a central region with coexisting pairing and charge correlations surrounded by a halo of staggered longitudinal spin correlations around $n \sim 1-m$, at $n=0.875$. These correlations provide experimentally accessible signatures of finite-momentum pairing beyond pair correlations. The Holstein model also offers a promising route toward quantum simulation, with proposals based on dressed Rydberg atoms and optical lattices \cite{Hague2020}, as well as trapped-ion platforms \cite{Mezzacapo2012,Stojanovic2012}. Our results therefore motivate experimental studies of spin-imbalanced systems, where the predicted pairing and correlation textures could be probed. Finally, in the anti-adiabatic limit, a particle-hole transformation maps our work onto a doped, spin-imbalanced repulsive Hubbard model, motivating further exploration of its rich interplay of pairing, charge, and spin correlations in a comparatively less studied (i.e.~non-zero magnetization) regime.

\vskip0.05in \noindent
\textit{Acknowledgments.}  
CF thanks Shiwei Zhang, Martin Zwierlein, and Yuan-Yao He for helpful discussions on FFLO physics, and Benjamin Cohen-Stead for assistance with the use of the SmoQy package.
RTS is supported by the grant DOE DE-SC0014671 funded by
the U.S.~Department of Energy, Office of Science. Calculations in this work are performed using SmoQy~\cite{Cohen-Stead2024}.

\bibliography{holsteinfflo}

\clearpage

\centerline{\bf Supplemental Materials}
\vskip0.10in

\vskip0.10in \noindent
These Supplemental Materials provide additional results supporting the main text.
We first present the sign average over the parameter regimes studied here (S1). We then discuss finite lattice size effects (S2), and finally present results at additional densities (S3).

\setcounter{figure}{0}
\renewcommand{\thefigure}{S\arabic{figure}}

 \begin{figure}[h]
  \includegraphics[width=1\columnwidth]{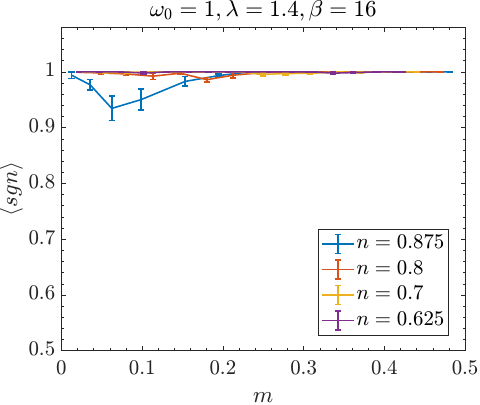}
 \caption{Average Monte Carlo sign as a function of magnetization $m$ for four densities, as indicated in the legend, at $\beta=16$.
 }
  \label{fig:sign_avg}
  \end{figure}

   \begin{figure}[t]
  \includegraphics[width=1\columnwidth]{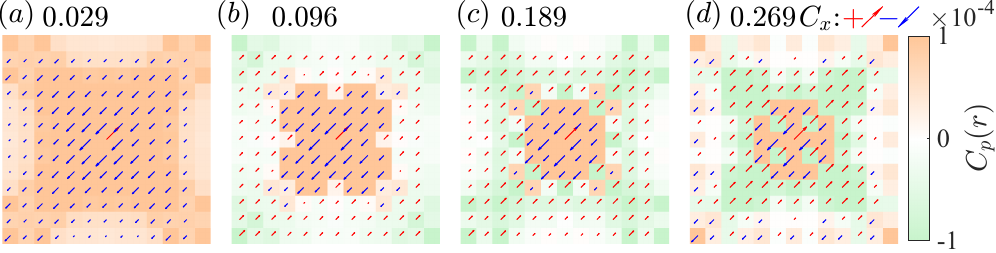}
    \includegraphics[width=1\columnwidth]{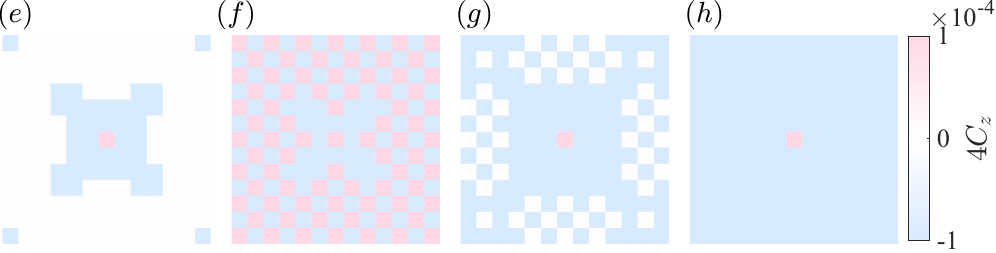}
      \includegraphics[width=1\columnwidth]{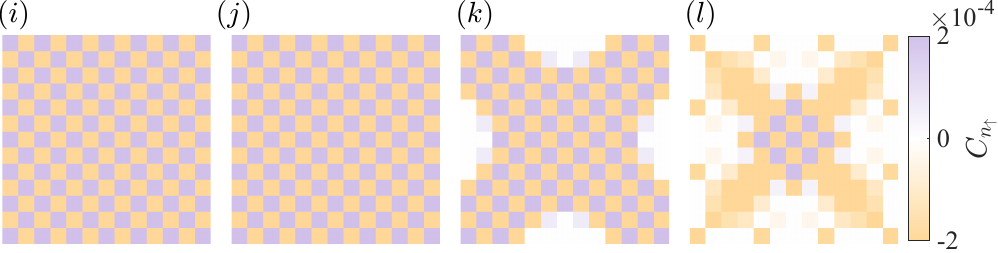}
 \caption{The same as Fig.~\ref{fig:lambda14_den0875_Sx_pairing} except that the calculation is on a $12 \times 12 $ lattice.
 }
  \label{fig:lambda14_den0875_Sx_pairing_L12}
  \end{figure}

   \begin{figure}[t!]
  \includegraphics[width=1\columnwidth]{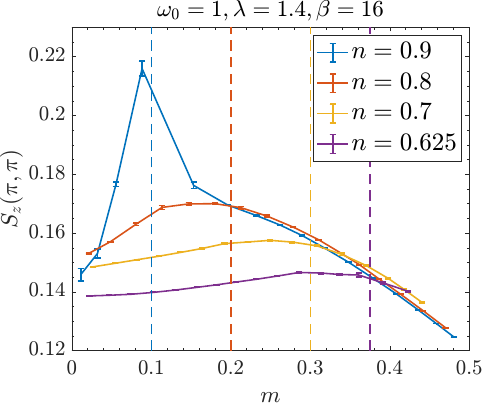}
 \caption{Spin-Z structure factor $S_z(\pi,\pi)$ as a function of magnetization $m$ for four different densities as shown in the legend. Dashed lines denote the position of $m=1-n$, around the peak position of different curves for different densities.
 }
  \label{fig:Sz_vs_m_diff_n}
  \end{figure}

   \begin{figure}[t!]
  \includegraphics[width=1\columnwidth]{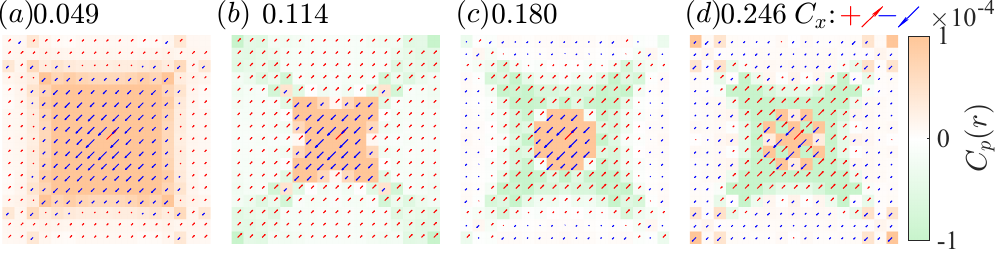}
    \includegraphics[width=1\columnwidth]{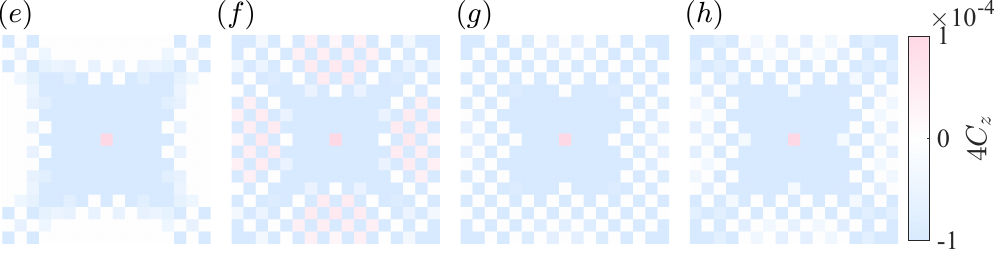}

 \caption{Similar to Fig.~\ref{fig:lambda14_den0875_Sx_pairing} but for density $n=0.8$.
 }
  \label{fig:lambda14_den08_Sx_pairing}
  \end{figure}

   \begin{figure}[t!]
  \includegraphics[width=1\columnwidth]{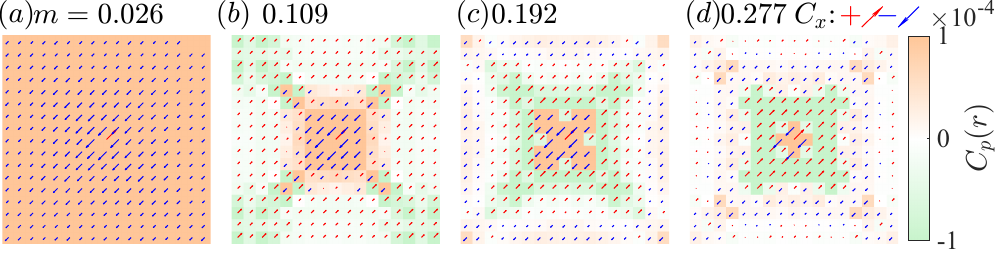}
    \includegraphics[width=1\columnwidth]{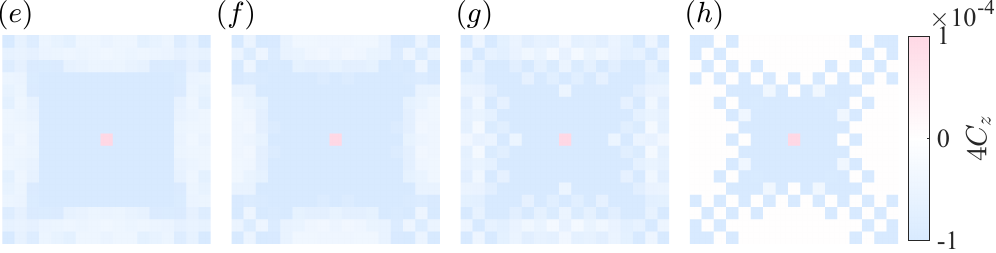}
 \caption{Similar to Fig.~\ref{fig:lambda14_den0875_Sx_pairing} but for density $n=0.7$.
 }
  \label{fig:lambda14_den07_Sx_pairing}
  \end{figure}
\vskip0.10in \noindent
\textit{S1.~The sign average-} 
Figure~\ref{fig:sign_avg} shows the average Monte Carlo sign as a function of magnetization $m$ for four representative densities at the lowest temperature studied here, $\beta=16$.
Across the parameter regimes considered in the main text, the average sign remains above $0.9$, indicating that the sign problem is essentially negligible and does not impose a practical limitation on the simulations.
The simulations can therefore reach the low-temperature regime with $\sim10^4 - 10^5$ sweeps, without the exponential sampling cost that typically accompanies a severe sign problem.
This is in sharp contrast to the spin-imbalanced attractive Hubbard model, where the sign problem becomes severe at low temperatures.
For example, for $U=-4$, the average sign can fall below $0.1$, and even below $0.01$, for $\beta\gtrsim5$, rendering unbiased DQMC calculations at such temperatures practically infeasible.
The exceptionally mild sign problem of the Holstein model thus enables unbiased access to the low-temperature regime where the finite-momentum pairing reported in the main text becomes pronounced.

\vskip0.10in \noindent
\textit{S2.~Finite lattice size effects-} 
To assess finite-size effects, we compare the results on a $12\times12$ lattice with those on the $16\times16$ lattice presented in Fig.~\ref{fig:lambda14_den0875_Sx_pairing} of the main text.
The corresponding results are shown in Fig.~\ref{fig:lambda14_den0875_Sx_pairing_L12}.
The spatial patterns on the smaller lattice closely resemble those in the central region of the $16\times16$ system, indicating that the main spatial signatures are insensitive to the system size.
In panels (a)--(d), increasing magnetization leads to increasingly pronounced sign-changing pairing correlations, with a shorter modulation wavelength at larger $m$, signaling finite-momentum pairing.
At low temperature $\beta=16$, the transverse-spin correlations exhibit the same modulation wavelength and node positions as the pairing correlations, apart from the onsite correlation.
In panels (e)--(h), staggered longitudinal-spin correlations are present for both $m=0.096$ and $m=0.189$, around the value $m=1-n=0.125$.
In panels (i)--(l), checkerboard charge correlations extend across the system at high density and small magnetization, $n=0.875$ and $m\lesssim0.1$.
With increasing magnetization $m=0.189$, the spin-up density correlations near the four edges of the system become strongly suppressed, foreshadowing the $\pi$-phase-shifted charge pattern seen on the larger $16\times16$ system in Fig.~\ref{fig:lambda14_den0875_Sx_pairing}. At larger magnetization, $m=0.269$, the checkerboard charge correlations are largely confined to the sites surrounding the central region. Overall, these comparisons confirm that the spatial patterns reported in the main text are robust against the system size.

\vskip0.10in \noindent
\textit{S3.~Results at other densities-} 
As discussed in the main text, the FFLO state with a staggered $S_z$ halo occurs predominantly around $m\sim1-n$, corresponding to a half-filled majority-spin, $n_{\uparrow}=0.5$. Figure~\ref{fig:Sz_vs_m_diff_n} shows the spin-Z structure factor as a function of magnetization for several densities. In addition to the $n=0.875$ case discussed in the main text, the enhancement of staggered spin correlations is consistently centered around $m\sim1-n$ for $n=0.9$, $0.8$, $0.7$, and $0.625$. The feature is relatively sharp at high densities, such as $n=0.9$, but becomes progressively broader for $n\lesssim0.8$.

Additional spatial results for $n=0.8$ and $n=0.7$ are shown in Figs.~\ref{fig:lambda14_den08_Sx_pairing} and \ref{fig:lambda14_den07_Sx_pairing}, respectively.
The finite-momentum pairing correlations retain the same qualitative behavior as at $n=0.875$: their modulation wavelength decreases with increasing magnetization, while the transverse-spin correlations exhibit the same modulation wavelength and node positions as the pairing correlations.
Near $m\sim1-n$, staggered longitudinal-spin correlations again develop around the finite-momentum-paired region, where the pairing correlations are weaker. Overall, the checkerboard charge correlations tend to weaken and become more spatially confined with decreasing density and increasing magnetization. At $n=0.8$ and $n=0.7$, charge patterns with a $\pi$-phase shift remain visible at small magnetization, consistent with related phenomena reported in the doped spin-balanced Holstein model \cite{Bradley2021}. A systematic investigation of the charge textures over a broader range of densities, magnetizations, and temperatures is beyond the scope of the present work and is left for future study.

\end{document}